\documentclass[10pt,letterpaper]{article}
\usepackage{opex3}

\usepackage{here}

\newcommand\pictc[5]{\begin{figure}[t,h]
                   \centerline{\vspace{-3mm}
                   \includegraphics[width=#1\columnwidth]{#3}}
               \protect\caption{\protect\label{fig:#4} #5}
                \end{figure}            }

\newcommand\pict[4][0.5]{\pictc{#1}{!tb}{#2}{#3}{#4}}
\newcommand\rpict[1]{\ref{fig:#1}}

\begin{document}
\begin{sloppy}

\title{Focusing and correlation properties of white-light optical vortices}

\author{Vladlen Shvedov$^{1,2}$, Wieslaw Krolikowski$^3$, Alexander Volyar$^2$, Dragomir N. Neshev$^1$, Anton S. Desyatnikov$^1$, and Yuri S. Kivshar$^1$}

\address{$^1$Nonlinear Physics Center, Research School of Physical Sciences and Engineering, Australian National University, Canberra ACT 0200, Australia \\
$^2$ Department of Physics, Taurida National University, Simferopol 95007 Crimea, Ukraine\\ 
$^3$Laser Physics Center, Research School of Physical Sciences and Engineering, Australian National University, Canberra ACT 0200, Australia}

\email{asd124@rsphysse.anu.edu.au}%% email address is required
\homepage{http://www.rsphysse.anu.edu.au/nonlinear} %% author's URL, if desired

\begin{abstract}
We generate double-charge white-light optical vortices by sending a circularly polarized partially incoherent light through an uniaxial crystal. We show that the generated polichromatic vortices are structurally stable, and their correlation properties can be altered by the beam focusing, resulting in changes of the vortex core visibility.
\end{abstract}


%--------------------------------------------------------------------------
\section{Introduction}
%--------------------------------------------------------------------------

Vortices are fundamental objects in the physics of waves, and they can be found in different types of coherent systems. A vortex has the amplitude vanishing at its center and a well-defined $2\pi n$ phase ramp associated with the circulation of momentum around the helix axis~\cite{berry}. In optics, vortices are associated with phase dislocations (or phase singularities) carried by optical beams~\cite{grover,soskin}. Last few years have seen a resurgence of interest in the study of optical vortices~\cite{list, Desyatnikov:2005:ProgressOptics}, owing in part to readily available computer-generated holograms and other techniques for creating phase singularities in coherent laser beams.

However, if a light beam is polychromatic and/or partially incoherent, the phase front topology of the vortex-carrying beam is not well defined, and statistical description is required to quantify the beam structure~\cite{Wolf}. In the incoherent limit neither the helical phase nor the characteristic zero intensity at the vortex core can be observed. Nevertheless, several recent theoretical and experimental studies have shed light on the question how phase singularities can be unveiled in partially coherent light fields~\cite{berry2, berry3, gbur, pono0, pono, Schouten:2003-968:OL, padgett, Palacios:2004-143905:PRL, volyar,maleev, soskin2,motzek, angel}. In particular, Palacios {\em et al.}~\cite{Palacios:2004-143905:PRL} used both experimental and numerical techniques to explore how a beam transmitted through a vortex phase mask changes as the transverse coherence length at the input of the mask varies. Assuming a quasi-monochromatic, statistically stationary light source and ignoring temporal coherence effects, they demonstrated that robust attributes of the vortex remain in the beam, most prominently in the form of a ring dislocation in the vortex cross-correlation function. The existence of a stable ring structure in the cross-correlation function has been shown in the effect of vortex stabilization in a self-focusing nonlinear medium~\cite{motzek}.

In this paper, we generate experimentally double-charge white-light optical vortex with different degrees of spatial coherence. Instead of using a holographic phase mask for imprinting the vortex phase, we employ a generation technique suggested by the group of Volyar~\cite{volyar2,volyar3,volyar4}. With this technique, a circularly polarized polychromatic light beam is tightly focused inside an uniaxial crystal and subsequently filtered by a polarizer, giving rise to generation of an optical vortex of double charge with zero chromatic dispersion. First, we reveal that the generated double-charge polychromatic vortices are structurally stable during their long-distance propagation, meaning that all spectral components remain co-axial and do not exhibit spectral anomalies near the vortex core. The vortex visibility in the transverse cross-section of the light beam, however, depends strongly on the correlation properties of the white-light source. Second, we study the process of focusing of such double-charge white-light vortices and reveal that the vortex correlation properties can be altered by the beam focusing. In particular, we demonstrate that a vortex generated in a spatially partially-coherent beam can become invisible at the focal point for relatively weak incoherence of the input beam. In contrary, an initially invisible vortex imprinted into a highly incoherent white-light beam may reappear again after the focus. The effect of recovery of the near-field vortex image by the lens implies that the information about the vortex can be effectively transfered from the cross-correlation function to its intensity distribution (auto-correlation function).

\section{Generation of polychromatic partially-coherent double-charge vortices}

%--------------------------------------------------------------------------
\pict[0.95]{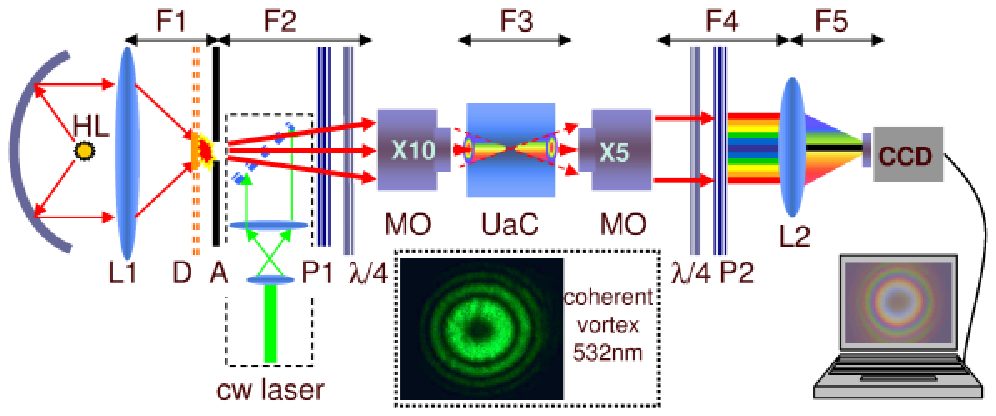}{exp_setup}{Experimental setup for the generation of partially incoherent double-charge vortices; HL - halogen white-light source (50 W), L - lenses, D - diffuser, A - aperture, P - polarizers, $\lambda/4$ - quarter-wave plate (532 nm), MO - microscope objectives, UaC- uniaxial crystal.}
%--------------------------------------------------------------------------

The experimental setup for generating white-light vortices is shown in Fig.~\rpict{exp_setup}. A collimated light beam generated by a halogen white-light lamp (HL) (with the angular divergence 13$^{\circ}$) is focused by the lens (L1, focal length $F_1 = 120$~mm) onto a diffuser (D), introducing a random-varying phase across the beam transverse section in the focal plane of L1. The magnitude of the solid angle of the light after the diffuser determines the degree of spatial coherence of the generated polychromatic light. This angle can be controlled by the size of the aperture (A) variable from the diameter $D=1$~mm to $13$~mm. Alternatively, the solid angle can be changed by adjusting the position of the aperture away from the diffuser. After the aperture, the light passes through a polarizer (P1) and $\lambda/4$ plate attaining circular polarization. Subsequently, the light beam is focused by a microscope objective (MO) with an aperture of 8~mm onto an uniaxial sapphire crystal (UaC) with the optical axes oriented along the beam propagation. The distance between the effective source and the microscope objective remains unchanged ($F_2=110$~mm), so that the degree of spatial coherence of the polychromatic light is solely defined by the size of the aperture. After the crystal the light is collimated by a second microscope objective and passes through a second $\lambda/4$ plate and a polarizer. The crystal changes the degree of ellipticity of polarization of the light beam. Then, by rotating the second polarizer (P2) we can extract a linearly polarized double-charge optical vortex as earlier predicted by Volyar {\em et al.}~\cite{volyar2,volyar3,volyar4}. The transverse intensity distribution of the vortex is registered by a colour CCD camera that can be translated along the beam propagation axis. This arrangement allows to scan the transverse structure of the polychromatic field during its propagation.

The implementation of our method for vortex generation has several advantages over the commonly used methods based on computer generated holograms and spiral phase plates. Firstly, the method is intrinsically free from chromatic dispersion and avoids anomalous spectral behavior near the vortex core~\cite{berry2,padgett,soskin2,Gbur:2002-13901:PRL}. Compensation of the chromatic dispersion can also be achieved with prism~\cite{padgett} or grating compensators~\cite{Bezuhanov:2004-1942:OL}, however our method provides high conversion efficiency into the vortex mode. Effects of topological dispersion for the different spectral components, as present in the vortex generation from a spiral phase plate~\cite{Swartzlander:2004-93901:PRL}, are also absent.

To control the generation and propagation of an optical vortex we employ an additional laser beam and combined it with the polychromatic light onto a beam splitter. The laser light ($\lambda = 532$~nm) was passed through a system of lenses in such a way that its divergence corresponded to that of the polychromatic light. Blocking one of the beams we can observe the field distribution in monochromatic or white light. The phase dislocations in both cases coincide, thus the use of the additional laser light allows us to compare the vortex propagation in coherent and incoherent cases. As shown in the inset of Fig.~\rpict{exp_setup}, the vortex generated in the green laser light posses a distinct dark core in the centre, followed by a dark ring-type phase dislocation, similar to the Airy ring patterns in the diffraction of a beam from a finite aperture.

%--------------------------------------------------------------------------
\pict[0.95]{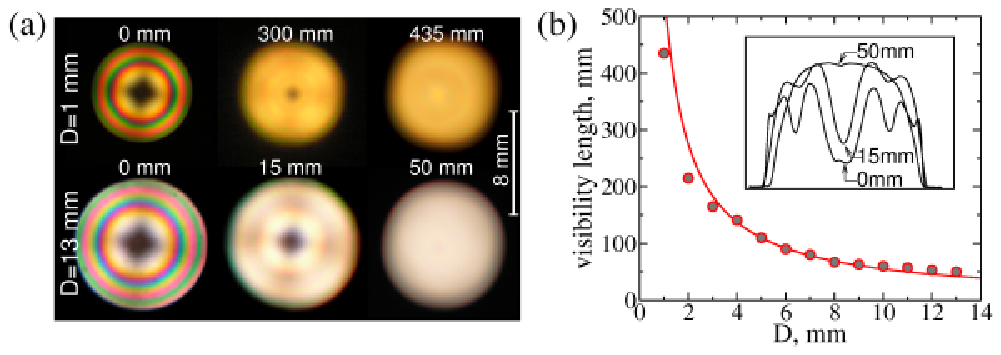}{visibility}{(a) Color images of the vortex beam at different distances after the crystal for two different apertures: $D=1$~mm (upper row) and $D=13$~mm (lower row). (b) Visibility length as a function of aperture size $D$. Solid line is $\sim1/D$ fit to the experimental points. Inset: transverse beam profiles for the case of low coherence.}
%--------------------------------------------------------------------------

In our first experiments, we varied the beam spatial coherence and revealed a number of specific features in the generation and evolution of a polychromatic optical vortex. The intensity of polychromatic vortices generated at two different degrees of spatial coherence are shown in Fig.~\rpict{visibility}(a, left), and we also tested interferometrically the imprinted double-charge phase structure. The vortex visibility in the polychromatic beam depends strongly on the spatial coherence, as was noticed in earlier studies (see, e.g., Ref.~\cite{Palacios:2004-143905:PRL}). As a result, the intensity of strongly incoherent light shows no sign of the imprinted vortex after some propagation, and the characteristic vortex structure is ``washed out'' completely. Figure~\rpict{visibility} shows three examples of the vortex beams transverse section, for two different degrees of spatial coherence (different aperture sizes). In the case of higher coherence (upper row), the vortex becomes invisible after propagating 435~mm from the crystal, while in the case of low coherence, this distance is much shorter, 50~mm. This behavior can be intuitively explained by the overlap of different spatial mutually incoherent components. A characteristic parameter of the process of vortex ``wash-out'' is the visibility length. It is determined by the propagation distance at which the vortex core completely disappears and the intensity distribution resembles a flat-top Gaussian beam, as shown in Fig.~\rpict{visibility}(b-inset) with profiles of total intensity corresponding to the case of lower coherence. The experimentally measured visibility length, shown in Fig.~\rpict{visibility}(b), is well approximated by a $\sim 1/D$ dependence (solid line).

The presence of a vortex in a partially coherent beam can be characterized by the mutual coherence function
\begin{equation}
\label{eq1}
\Gamma(\mathbf{r}_1,\mathbf{r}_2;z) =
\left<E^*(\mathbf{r}_2,z,t)E(\mathbf{r}_1,z,t)\right>,
\end{equation}
where the brackets stand for averaging over the net field, so that the intensity is given by the auto-correlation function, $I(\mathbf{r};z)=\Gamma(\mathbf{r},\mathbf{r};z)$. The intensity of a fully coherent vortex preserves the characteristic zero point at the origin during the propagation, while the intensity of spatially incoherent vortex beam is blurred in the far-field. Palacios {\em et al.}~\cite{Palacios:2004-143905:PRL} demonstrated that the presence of a phase singularity corresponds to ring-type dislocation in the cross-correlation function $\Gamma(\mathbf{r},-\mathbf{r};z)$ of an incoherent vortex beam. The radius of the ring dislocation, $R$, grows with decreasing of the relative coherence length, $\sigma=L_c/w$, where $L_c$ is the transverse coherence length of a source and $w$ is the beam size~\cite{Palacios:2004-143905:PRL,maleev}. Subsequently, the visibility of the partially coherent vortex can be associated with the radius $R$, so that $R=0$ for fully coherent beam. In particular, the visibility of vortex core vanishes if the correlation dislocation radius exceeds some value $R>R_c$. Using the estimations for the radius $R(z)$~\cite{Palacios:2004-143905:PRL} we find that indeed the visibility length, $z_c=k(2\pi)^{-1/2}R_cL_c$, is inversely proportional to the source transverse size $D$, measuring the coherence length, $L_c\sim 1/D$. These estimations agree well with experimentally measured dependence in Fig.~\rpict{visibility}(b).

An important characteristic of the propagation of the polychromatic vortex beam is that all spectral components propagate coaxially near the vortex core and this region has typical white colour (depending on the emission spectrum of halogen lamp). In contrast, the Airy rings surrounding the vortex, representing edge-type phase dislocation, exhibit strong coloring in the form of rainbow rings. Such behavior is to be expected as theoretically predicted in Ref.~\cite{Ponomarenko:15-1211:OL} and experimentally demonstrated in Ref.~\cite{Popescu:2002-183902:PRL}. In the later stages of propagation the rainbow rings are also wiped-out due to the large spatial incoherence of the beam.

\section{Focusing of a white-light vortex}

In order to test the structural stability of the generated white-light vortices we performed experiments on focusing of the vortex beam by a lens. Here, we demonstrate that even after the process of vortex wash-out, its near-filed intensity structure can be recovered by a simple focusing of the polychromatic singular beam. To show this, we introduce an additional lens L2 into the beam path (Fig.~\rpict{exp_setup}). The distance from the collimator to the lens $F_4$ is varied depending on the particular point of free-space vortex evolution we aim to study. In addition, at a fix position of the lens L2 we vary the aperture size changing the coherence of the effective source. 

We conducted two different experiments for the focusing of the spatially incoherent vortex, and the results are summarized in Fig.~\rpict{focusing1} and Fig.~\rpict{focusing2}. In the first case (see Fig.~\rpict{focusing1}), we fixed the size of the aperture to $D=1$~mm, and placed a lens with a focal length $f=60$~mm at a distance $F_4=100$~mm from the collimator.  The beam intensity distribution just before the lens is shown on the left side of Fig.~\rpict{focusing1}. The dark core of the vortex is clearly seen in the transverse structure of the beam for any point between the crystal and the lens. However, near the focal point of the lens the vortex is ``washed out'', it becomes invisible in the transverse beam profile, as is seen in Fig.~\rpict{focusing1} (image at 0 mm). Behind the focal point the vortex reappears again, being almost identical to the input vortex.

%--------------------------------------------------------------------------
\pict[0.99]{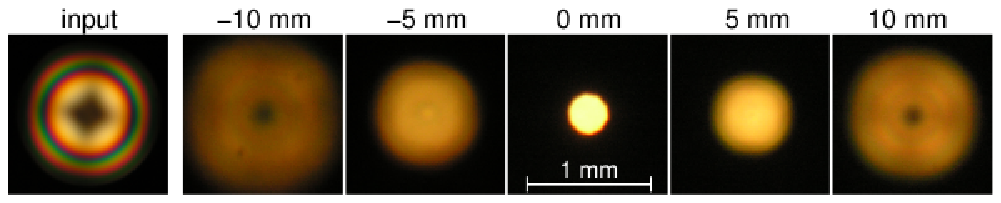}{focusing1}{Focusing of the polychromatic vortex beam: $f=60$~mm, $D=1$~mm. Left -- beam before the lens; right -- intensity distribution after the lens for different positions relative to the focal point.}
%--------------------------------------------------------------------------

%--------------------------------------------------------------------------
\pict[0.99]{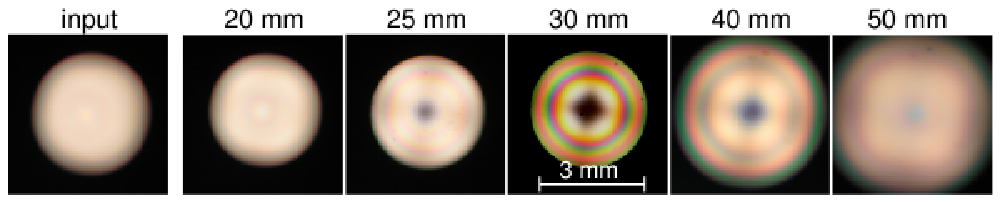}{focusing2}{The same as in Fig.~\rpict{focusing1} for $f=35$~mm and $D=13$~mm. Left -- beam before the lens; right -- intensity distribution for different distances behind the focal point.}
%--------------------------------------------------------------------------

In the second case (see Fig.~\rpict{focusing2}), we set the size of the aperture to a larger value $D=13$~mm, that defines larger incoherence of the optical beam, and placed the lens L2 of a focal length $f=35$~mm at a distance $F_4=100$~mm from the collimator.  After the propagation of 100~mm the beam intensity is homogeneous in front of the lens and no vortex structure is visible [see Fig.~\rpict{focusing2} (input)], indicating that the beam is highly spatially incoherent. Near the focal point the vortex is not yet visible (it is hidden~\cite{Gbur:2004-S239:JOptA}), but it starts to reappear after the focus and it is fully recovered at the image plane, as shown in Fig.~\rpict{focusing2} at 30~mm, where the distances are measured from the focal plane. This effect indicates a simple mean to recover the hidden information about the phase singularities~\cite{Gbur:2004-S239:JOptA}, otherwise only possible through the analysis of the cross-correlation function~\cite{Palacios:2004-143905:PRL}. We expect that the analysis of the beam cross-correlation function~\cite{Palacios:2004-143905:PRL, Gbur:2002-2097:JOSAA} can be applied to the case of a double-charge vortex discussed here and it may reveal interesting features to explain the effect of the vortex transformation observed experimentally.

%--------------------------------------------------------------------------
\section{Conclusions}
%--------------------------------------------------------------------------
We generated double-charge white-light optical vortices by sending a circularly polarized polychromatic light beam through an uniaxial crystal and showed that these vortices are structurally stable in propagation. By measuring the vortex visibility length for different degrees of coherence of the source, we explored the correlation properties of such polychromatic vortices. We demonstrated that the vortex visibility can be altered by the beam focusing, so that vortex core disappears near the focal region, while an initially invisible vortex imprinted into a white-light beam will become visible after the focal point. These results can be interpreted as the exchange of the information about the vortex between its intensity distribution and the cross-correlation function carrying a robust ring-type dislocation.

\section{Acknowledgements}
The authors thank G. Swartzlander and S. Ponomarenko for useful discussions and acknowledge a support of the Australian Research Council. V. Shvedov and A. Volyar thank Nonlinear Physics Centre for a warm hospitality during their stay at the Australian National University.

%==========================================================================
\end{sloppy}
\end{document}